\documentclass[12pt]{article}
\usepackage[dvips]{graphicx}
\begin{document}
\sf
\begin{center}
   \vskip 2em
{\LARGE \sf   Axially symmetric membranes with polar tethers}
%\vskip1em {\LARGE \sf non-vanishing Noether Charges as polar
%tethers}
\vskip 3em
 {\large \sf  Pavel Castro-Villarreal and Jemal Guven \\[2em]}
\em{ Instituto de Ciencias Nucleares,
 Universidad Nacional Aut\'onoma de M\'exico\\
 Apdo. Postal 70-543, 04510 M\'exico, DF, MEXICO\\[1em]
}
\end{center}
 \vskip 1em

\begin{abstract}
Axially symmetric equilibrium configurations of the conformally
invariant Willmore energy are shown to satisfy an equation that is
two orders lower in derivatives of the embedding functions than the
equilibrium shape equation, not one as would be expected on the
basis of axial symmetry. Modulo a translation along the axis, this
equation involves a single free parameter $c$.
%An isothermal parametrization of the
%surface geometry which exploits the conformal symmetry is used to
%derive this equation. In this parametrization, the Euler-Lagrange
%equation is cast as a sine-Gordon equation with a tunable non-local
%source. This source originates in constraints associated with the
%geometry not captured by the sine-Gordon equation itself.
%This modification alters profoundly the properties of the equation.
%This equation possesses  a first integral; remarkably,  it is two
%Reparametrization invariance places a constraint on the value of the
%first integral. The properties of solutions with spherical topology are described.
If $c\ne 0$,  a geometry with spherical topology will possess
curvature singularities at its poles. The physical origin of the
singularity is identified by examining the Noether charge associated
with the translational invariance of the energy; it is consistent
with an external axial force acting at the poles. A one-parameter family of
exact solutions displaying a discocyte to stomatocyte transition is
described.
\end{abstract}
%\today
 \vskip 3em

\section{\sf Introduction}

The course-grained description of membranes and interfaces often
involves purely geometrical degrees of freedom. The energy
associated with tension is proportional to area. The positive
definite invariant,
\begin{equation}
H = {1\over 2} \int dA\,  K_{ab} K^{ab}\,, \label{eq:Hdef0}
\end{equation}
quadratic  in the surface extrinsic curvature $K_{ab}$, is a measure
of the energy associated with bending \cite{Willmore} (the notation
we adopt is defined in an appendix). This energy has the remarkable
property that it is invariant with respect to conformal
transformations of the ambient three-dimensional space. It plays a prominent role in
the physics of soft matter, notably in the description of fluid
membranes and liquid crystals (see, for example, \cite{Nel,Kleman}).
Physically realistic models will, of course, involve additions to $H$ which
break the conformal symmetry
\cite{canhamHelfrich,svetina89,seifert}: there may be tension and an
osmotic pressure associated, respectively, with global constraints
on the area and on the volume; there also may be an asymmetry
reflected in an energy linear in curvature. There are, however,
important lessons we can learn from the simple description of the surface
provided by Eq.(\ref{eq:Hdef0}). For it is still possible to constrain
the geometry using local constraints which break the conformal
invariance of Eq.(\ref{eq:Hdef0}) only at isolated points.

The Euler-Lagrange equation corresponding to Eq.(\ref{eq:Hdef0}) is
given by ($\nabla^2$ is the surface Laplacian)
\begin{equation}
  -\, \nabla^2 K +
{1\over 2} K (K^2 - 2 K_{ab} K^{ab} ) =0\,.
 \label{eq:elch}
\end{equation}
The few known analytical solutions of Eq.(\ref{eq:elch}) are easily
identified by inspection: these are spheres with $K_{ab}= g_{ab}
K/2$, as well as the minimal surfaces with $K=0$. In these
geometries the membrane is free of stress. There is also the
well-known Clifford torus  with ratio of wheel to tube radius of
$\sqrt{2}$. In practice, one relies overwhelmingly on numerical
analysis to map out the configuration space (some recent developments are
described in \cite{Wor,Du,Zit}). On the other hand, in the early nineties
Seifert, Julicher, Lipowsky and others pointed out features of solutions
lying beyond the reach of numerical analysis that are direct consequences of
the conformal symmetry of the energy \cite{Seif1}.  After a brief flurry of activity
the subject waned. The goal of this paper will be
to underscore the point that the conformal symmetry of
Eq.(\ref{eq:Hdef0}) merits reexamination.

We will first show that, modulo a translation parallel to its axis,
all axially symmetric equilibrium shapes satisfy the following
equation
\begin{equation}
{R^2\over 2}\left( C_\perp^2 - C_\|^2\right)  - c\, {\bf n}\cdot
{\bf X}
 = 0 \,.
\label{eq:quadOmega1}
\end{equation}
Here $C_\perp$ and $C_\|$ are the principal curvatures along the
meridian and along the parallel, $R$ is the polar radius, and ${\bf
n}\cdot {\bf X}$ is the support function. This equation is two
orders lower in derivatives of the embedding functions ${\bf X}$
than Eq.(\ref{eq:elch}), not one
derivative, as would have been expected on the basis of axial
symmetry. It involves a single free parameter $c$.
%whose role is to
%couple the extrinsic geometry (as characterized by the rotation
%angle along the meridian $\Omega$) to the intrinsic geometry.
%The origin of the source term is a pair of constraints associated
%with the geometry.

The conformal invariance of the energy may be exploited to construct
a one-parameter family of solutions of Eq.(\ref{eq:quadOmega1}).
This symmetry implies that an equilibrium geometry maps to another equilibrium
geometry under conformal transformation.
A catenoid is a minimum surface and thus an
equilibrium. By inverting the catenoid  in points along its axis
a sequence of deflated compact axially symmetric equilibrium geometries
displaying a discocyte to stomatocyte transition
is generated.

It is also possible to identify important properties of solutions of
Eq.(\ref{eq:quadOmega1}) without having to solve it explicitly. The only
solution with spherical topology that is regular at its poles is the
round sphere with $c=0$. If $c\ne 0$, however, all solutions with
this topology exhibit curvature singularities at their poles.
In general, this parameter determines the strength of the
singularity. The physical interpretation of $c$
is as an external force pulling the poles of the membrane
together. This identification is based on the existence of a
conserved stress tensor on the membrane associated with
translational invariance. The forces are the Noether charges
associated with this symmetry \cite{Stress}. The potential relevance
of these solutions to modeling tethered membranes will be discussed.

\section{\sf Axially symmetric equilibrium shapes}

%change here
We begin by deriving Eq.(\ref{eq:quadOmega1}). Our approach will be
to adapt the variational principle to a parametrization that
exploits both the conformal symmetry of the energy as well as the
axial symmetry of the configuration. Technically, this is simple.
However, it is also all too easy to overlook constraints or to lose
sight of the dependence on the geometry implicit in the
parametrization. It is therefore useful to go through the details
carefully. The equation itself will be independent of the
parametrization.

An axially symmetric surface is described in terms of cylindrical
polar coordinates $(\rho,z,\phi)$ on $R^3$. The axis of symmetry
coincides with the $z$-axis. There are two parametrizations of the
surface tailored to this symmetry that we will find useful. The more
straightforward of the two involves arc-length $l$ along the
meridian and the angle $\phi$ along the parallels. The surface is
then described by two functions $\rho= R(l)$, and $z=Z(l)$; these
functional relationships are not independent; they satisfy the
constraint $R'{}^2 + Z'{}^2 =1$ implied by the parametrization,
where the prime denotes a derivative with respect to arc-length. The
line element on the surface is now given by
\begin{equation}
ds^2 = dl^2 + R(l)^2 d\phi^2\,. \label{eq:l1}
\end{equation}
An isothermal parametrization is obtained by replacing arc-length
$l$ by a parameter $u$ defined in such a  way that the line element
assumes the conformally flat form,
\begin{equation}
ds^2 = e^{2\sigma (u)}(du^2 + d\phi^2)\,. \label{eq:l2}
\end{equation}
Comparison of the two expressions (\ref{eq:l1}) and (\ref{eq:l2})
identifies $R = e^\sigma$, and connects the parameter $u$ to
arclength through the relation $\partial_u l = R$.

To characterize the extrinsic geometry, introduce the variable
$\Omega$ representing the angle which the outward normal to the
surface makes with the (positive) axis of symmetry; the tangent
along the meridian ${\bf l}$ is then given by $(R',Z',0)=
(\cos\Omega,-\sin \Omega,0)$. At any point on the surface, ${\bf l}$
together with the tangent along the parallel ${\bf t}$ define the
principal axes of the surface. The corresponding curvatures are
given respectively by
\begin{equation}
C_\perp = \Omega' = {\partial_u\Omega\over R}\,, \quad\quad
C_\parallel  = {\sin \Omega\over R}\,.
 \label{curvatures}
\end{equation}
As noted in \cite{BenSax},  in terms of isothermal coordinates, the
bending energy then assumes the deceptively simple form
\begin{equation}
H[\Omega] = 2\pi \, \int du \, {\cal H}_0\,,
\end{equation}
where
\begin{equation}
{\cal H}_0= {1\over 2}\, \left\{ (\partial_u\Omega )^2 + \sin^2
\Omega \right\}\,. \label{eq:calH0}
\end{equation}
In particular, $H$ appears to depend only on $\Omega$. Importantly,
$\sigma$ does not appear in  $H$, at least not explicitly. This is a
direct consequence of the global conformal invariance of the energy
\cite{Willmore}.

To determine the Euler-Lagrange equation for $\Omega$, it is
necessary to impose constraints on $\Omega$ to reflect the fact that
$\Omega$ parametrizes the normal vector. The variational principle
must therefore be consistent with the following local constraints,
\begin{equation}
\cos\Omega =  R^{-1} \partial_u R\,,\quad \sin\Omega = -  R^{-1}
\partial_u Z\,. \label{eq:cs}
\end{equation}
We thus introduce two local Lagrange multipliers $\lambda_R$ and
$\lambda_Z$ (as we see, one is generally not enough) and replace
${\cal H}_0$ by ${\cal H}= {\cal H}_0 + \Delta {\cal H}$ where
\begin{equation}
\Delta {\cal H} =
 \lambda_R (\cos\Omega -  R^{-1} \partial_u R)
+ \lambda_Z (\sin\Omega + R^{-1} \partial_u Z)\,. \label{eq:dcalH}
\end{equation}
The Euler-Lagrange equations for $R$ and $Z$ give
\begin{equation}
\partial_u \lambda_R + \sin\Omega \, \lambda_Z=0\,,
\end{equation}
and
\begin{equation}
\partial_u \lambda_Z - \cos\Omega \, \lambda_Z=0
\end{equation}
respectively. They do not involve ${\cal H}_0$. As such, they  are
easy to solve. The solution is
\begin{equation}
\lambda_Z= c\, R\,,\quad \lambda_R = c\, Z + d\,; \label{eq:lambdas}
\end{equation}
it involves two constants of integration, $c$ and $d$. Note that if
the latter constraint is overlooked, we miss the constant $c$. This
oversight is thus equivalent to setting $c=0$. The role of the constant $d$ is
trivial; it is clear that we can always set $d=0$ when $c\ne 0$ by
translating the geometry along the axis of symmetry.

The Euler-Lagrange equation for $\Omega$ is now given by
\begin{equation}
-\partial_u^2 \, \Omega + \sin\Omega \cos\Omega - \lambda_R \, \sin
\Omega + \lambda_Z\, \cos \Omega =0\,. \label{eq:elOmega}
\end{equation}
$\Omega$ couples to the intrinsic geometry through the two
multipliers $\lambda_R$ and $\lambda_Z$ given by
Eq.(\ref{eq:lambdas}). Using the constraints we can cast $\lambda_R$
and $\lambda_Z$ non-locally in terms of $\Omega$. This has the
appearance of a non-local generalization of the sine-Gordon
equation. However, as we will show this analogy is not very useful.
One can do much better.

If $c=0$, then $\lambda_Z$ vanishes and $\lambda_R$ is constant.
$\Omega$ then uncouples from the intrinsic geometry and one obtains
a (local) double sine-Gordon equation. Such an equation was obtained
recently \cite{BenSax1}. The non-local terms are removed. We will
show, however, that this limits the space of solutions
very severely: only spheres and
catenoids remain.

Remarkably, it is possible to integrate Eq.(\ref{eq:elOmega}) to
produce a `quadrature':
\begin{equation}
{1\over 2} (\partial_u\Omega)^2 - {1\over 2}\,\sin^2\Omega -
\lambda_R \cos\Omega - \lambda_Z \sin\Omega = E \,.
\label{eq:quadOmega}
\end{equation}
The technical reason why this is possible is because the two
derivative terms involving the multipliers cancel:
\begin{equation}
\cos \Omega \, \partial_u \lambda_R + \sin\Omega \, \partial_u
\lambda_Z =0\,.
\end{equation}
There is a deeper geometrical reason which will be developed  below.
The result is that a quadrature is obtained even when $c\ne 0$ so
that neither $\lambda_R$ nor $\lambda_Z$ is constant. Of course, in
this case, unlike the sine-Gordon equation, it will not generally be
possible to integrate the quadrature.

There remains to address a subtlety associated with
reparametrization invariance that has been carefully sidestepped so
far: specifically, in the variational principle, it is necessary to
take into account the fact that the isothermal parametrization is
not some arbitrary parametrization of the surface: it depends
implicitly on the surface geometry. Physical solutions will need to
be consistent with the constraint implied by this dependence. As a
consequence, the constant of integration $E$ entering
Eq.(\ref{eq:quadOmega}) must vanish.

To see this, note that if the interval of integration $u_f-u_i$ is
fixed, a global constraint
\begin{equation}
\int {dl \over R} = u_f-u_i \label{eq:1/R}
\end{equation}
is placed on the geometry --- a constraint which is clearly
unphysical. While it does respect scale invariance, this constraint
spoils the conformal invariance of the problem. Thus if the position
along the meridian is parametrized by the isothermal variable $u$,
$u$ itself must be allowed to vary freely at the endpoints within
the variational principle. Now, modulo Eqs.(\ref{eq:cs}),
(\ref{eq:lambdas}) and (\ref{eq:elOmega}), it is easily checked that
$\delta H= 2\pi( E(u_f)\, \delta u_f - E(u_i) \,\delta u_i)$, where
\begin{equation}
E(u) = {\partial {\cal H}\over \partial \, \partial_u \Omega} \, \partial_u\Omega +
 {\partial {\cal H}\over \partial\,\partial_u R} \, \partial_u R
 +{\partial {\cal H}\over \partial \,\partial_u Z} \, \partial_u Z - {\cal H} \,.
\end{equation}
Therefore, in a stationary configuration, $E(u)$ must vanish at the
endpoints. And because ${\cal H}$ does not depend explicitly on $u$,
$E$ must be constant. As a result, it is zero everywhere. This is,
of course, the constant appearing in the quadrature
(\ref{eq:quadOmega}). In the appendix, we will show that $E$ can be
identified as the global Lagrange multiplier associated with the
constraint (\ref{eq:1/R}). When $E$ vanishes, the unphysical constraint is
relaxed. This, of course, is not new: a completely analogous problem is
encountered in the parametrization of an axially symmetric geometry
by arc-length \cite{sources}; it is a subtlety that is perhaps more
familiar in the modeling of reparametrization invariant relativistic systems.
In this context it is often desirable to
introduce a parameter depending explicitly on the trajectory, such
as proper time, within the variational formulation of the dynamics
(see, for example, \cite{Farhi}). The issue is discussed in the
Hamiltonian setting in \cite{Hamo}.

The beauty of Eq.(\ref{eq:quadOmega}) is not its sine-Gordon form,
but its remarkably simple gauge invariant geometrical structure
which is manifest when it is expressed in terms of the principal
curvatures and the support function ${\bf n}\cdot {\bf X}$ to give
Eq.(\ref{eq:quadOmega1}).

In the next section, we will look at equilibrium from the point of
view of the stresses that exist within the membrane; this approach
will permit us to interpret the constant $c$ in terms of external
forces. Such an interpretation is not obvious within the simple
economical framework we have been using up to now. It will also
permit us to establish the consistency of Eq.(\ref{eq:quadOmega1})
with the corresponding first integral of the shape equation
(\ref{eq:elch}). What will become clear is that no single approach
is appropriate everywhere.

\section{\sf Stress and Noether charges}

The statement of equilibrium can be cast in terms of a conserved
stress tensor \cite{Stress}.  This framework does, unfortunately,
involve the introduction of a little extra formalism.

Using the notation defined in the appendix, the  stress in the
membrane is given by \cite{Stress}
\begin{equation}
{\bf f}^a = K (K^{ab}- {1\over 2} g^{ab} K) \,{\bf e}_b -\partial^a
K\, {\bf n}\,.
\label{eq:stress}
\end{equation}
A simple derivation is provided in \cite{auxil}) using
auxiliary variables. It has also been subject of a recent detailed
treatment \cite{LomMiao}. On the free surface of the membrane, ${\bf
f}^a$ is conserved in equilibrium so that
\begin{equation}
\nabla_a \,{\bf f}^a =0\,, \label{eq:conservf}
\end{equation}
where $\nabla_a$ is  the surface covariant derivative compatible
with $g_{ab}$. This description is manifestly invariant with respect to
reparametrizations of the surface. One can easily check that
Eq.(\ref{eq:conservf}) reproduces the shape equation
(\ref{eq:elch}). From a field theoretical point of view,
Eq.(\ref{eq:conservf}) is a direct consequence of the translational
invariance of the energy.  ${\bf f}^a$ is a Noether current. We will
now show that the constant $c$ appearing in Eq.(\ref{eq:quadOmega1})
is none other than the corresponding Noether charge.

In reference \cite{Stress}, the global statement of conservation
which follows from  Eq.(\ref{eq:conservf}) was discussed. The focus,
however, was on the behavior of an isolated regular topologically
spherical membrane subject to global internal constraints; thus the
boundary conditions that were admitted were not the most general.
They are certainly not the boundary conditions that are appropriate
here. For this reason, it will be useful to retrace the steps
involved in integrating the conservation law to accommodate
curvature singularities or non-trivial boundary conditions, and to
provide an interpretation.

Thus, consider any closed contour $\Gamma$ on the free surface of a
membrane. Using Stoke's theorem, the conservation law
Eq.(\ref{eq:conservf}) implies that the closed line integral
\begin{equation}
\oint ds \,\, l^a\, {\bf f}_a
\end{equation}
is a constant vector ${\bf F}$ along contours that are homotopically
equivalent to $\Gamma$ on this surface. Here $l^a$ is the normal to
$\Gamma$ tangent to the surface, and $ds$ is the element of
arclength. If the geometry is regular and the contour can be
contracted to a point this vector vanishes.  There are, however, various
possible obstructions: the membrane topology may be non-trivial, as
it is for a torus. There may also be a source of stress within the
region spanned by the contour. In the latter case the geometry may
be regular if the source is distributed over a finite area. A
idealized point source of stress, however, will be reflected in a
curvature singularity which is picked up by the line integral. ${\bf
F}$ has a clear interpretation as the net force acting on the region
\cite{MDG}.

In an axially symmetric geometry, this force can only be parallel to
the axis. We thus have (${\bf k}$ is a unit vector pointing along
the $z$ axis)
\begin{equation}
\oint ds \,\, l^a \,{\bf f}_a \cdot  {\bf k} =  2\pi c\,,
\label{eq:axicon}
\end{equation}
where $c$ is a constant.

In an axially symmetric geometry, it is particularly simple to
evaluate the integrand appearing in Eq.(\ref{eq:axicon}). Without
loss of generality, let the contour be a closed circle of fixed $Z$.
Then $l^a$ is the tangent to the meridian (${\bf l} = l^a{\bf e}_a$)
and $ds$ is the element of arclength along the parallel $ds = R
d\phi$. Now
\begin{equation}
l_a {\bf f}^a = {1\over 2} (C_\perp^2 - C_\parallel^2) \, {\bf l} -
(C_\perp + C_\parallel)' \, {\bf n}\,,
\end{equation}
so that
\begin{equation}
l_a {\bf f}^a \cdot {\bf k}  =   {1\over 2}\Big(
(\partial_u\Omega)^2 - \sin^2\Omega \Big)\, {\sin\Omega\over R^2} +
\, \Big(
\partial_u^2 \Omega  - \sin \Omega \cos\Omega\Big)\,
{\cos\Omega\over R^2}\,.
\end{equation}
The integrand appearing on the lhs of (\ref{eq:axicon}) does not
depend on $\phi$.

To establish consistency with the approach adopted in the previous
section, note that there is a linear combination of the
Euler-Lagrange equation (\ref{eq:elOmega}) and its `first integral'
(\ref{eq:quadOmega}) which permits the elimination of $\lambda_R$:
%(for later comparison we preserve $E\ne0$) :
\begin{equation}
{1\over2} \sin\Omega \, \Big( (\partial_u\Omega)^2 - \,\sin^2 \Omega
\Big) + \cos\Omega\, \Big(\partial_u^2 \Omega  - {1\over 2} \sin
2\Omega \Big) - \lambda_Z
%+ E \sin\Omega
= 0\,. \label{eq:sumelquad}
\end{equation}
It should now be clear that  equation (\ref{eq:sumelquad}) coincides
with the statement of conservation for the underlying stress tensor.
It is the first integral of the shape equation. Consistency requires
the identification of the two constants of integration. The
framework of a conservation law provides a remarkably direct
interpretation of the constant $c$ appearing in
Eq.(\ref{eq:quadOmega1}).

%In the appendix, we show how the shape equation itself may be recovered.
We note one important point: had one insisted on holding
on to a non-vanishing value of $E$ in equation (\ref{eq:sumelquad})
it would not have been consistent with (\ref{eq:axicon}).
Consistency requires both $E=0$ and the identification of the
constant of integration $c$ as anticipated. In the appendix, we also
show that the unphysical constraint on the range of the isothermal
parameter $u$ manifests itself in this framework through an additional
unphysical tangential stress.

It should be pointed out that we have not found a simple derivation
of Eq.(\ref{eq:quadOmega1}) that follows directly from
Eq.(\ref{eq:elch}). The first integral of the latter, described
above, is a weaker statement than Eq.(\ref{eq:quadOmega1}) for it
involves one extra derivative.

Of course, the existence of a first integral of the shape equation
for axially symmetric geometries is well known using alternative
approaches \cite{sources,firstintegral}. The appearance of a
constant of integration $c$ was also noted. As Zheng and Liu, among
others, have emphasized the physics does not depend on the
parametrization \cite{firstintegral}. However, the physical role of
this constant does not appear to have been appreciated. Posing the
problem in terms of a conservation law is not just a formal
exercise. What was not anticipated was that it would be possible to
do much better than Eq.(\ref{eq:sumelquad}). For, as we have seen,
we possess the stronger result, Eq.(\ref{eq:quadOmega1}) involving
not one but two integrations of Eq.(\ref{eq:elch}).

%Note that isothermal coordinates also suggest the
%compact form for this equation:
%\begin{equation}
%\partial_u \left( {I(\Omega,\partial_u\Omega)\over \cos\Omega}\right) =
% c\, {R\over \cos^2\Omega} \partial_u \Omega\,,
%\end{equation}
%where $I= (
%(\partial_u\Omega)^2 - \sin^2\Omega)/2$.
%The role of the constant $c$ in coupling $\Omega$ to $R$ is now explicit.

Before describing new solutions of Eq.(\ref{eq:quadOmega1}), it is a
useful exercise to identify various features of the stress tensor in
the familiar regular axially symmetric geometries: In the case of
the catenoid, a minimal surface with $C_\perp + C_\|=0$, it is
obvious from Eq.(\ref{eq:quadOmega1}) that $c=0$.
%Indeed, the only
%surfaces consistent with $c=0$ are patches of spheres or catenoids.
On the other hand,  the constant $c$ has a non-vanishing value on
the Clifford torus. Consider a circular torus characterized by two
radii, $a$ and $b$,
\begin{equation}
R = a + b \sin\Omega\,.
\end{equation}
Note that we now have $l= b \,\Omega$, so that $\partial_u \,\Omega
= R/b$.
%and $\partial^2_u \,\Omega = \cos\Omega R/b$.
It is now simple to check that $R$ is a solution of
Eq.(\ref{eq:quadOmega1}) if and only if $a=\sqrt{2} b$.  The value
of $c$ is fixed, $c= 1/b$ (and, of course $E=0$). In particular, $c$
does not vanish. The origin of the constant $c$ for the Clifford
torus lies clearly in its topology. Observe that whereas the stress
tensor itself, given by Eq.(\ref{eq:stress}) vanishes everywhere on
spheres and minimal surfaces, it does not on the Clifford torus.

\section{\sf New Solutions}

The conservation law provides the appropriate framework for
characterizing equilibrium geometries. If the parameter $c$
appearing in this equation does not vanish,  it is simple to see
that any topologically spherical solution of
Eq.(\ref{eq:quadOmega1}) must possess a curvature singularity at its
poles. The strength of this singularity is proportional to $c$.
Furthermore, the round sphere is the only non-singular solution of
Eq.(\ref{eq:quadOmega1}) with this topology. For if the geometry at
either pole is regular and lies on the free membrane, then the
contour determining $c$ can be contracted onto that pole. The
constant must therefore vanish: $c=0$. The boundary condition at the
pole, $\partial_u \Omega =0=\sin\Omega$, then implies that $d=0$ in
Eq.(\ref{eq:quadOmega}). As a consequence $C_\|=C_\perp$ on the free
surface. It is thus necessarily a part of a sphere.

The conformal invariance of the energy permits us to identify a
one-parameter family of solutions of Eq.(\ref{eq:quadOmega}). We
note that, under inversion in the origin,
\begin{equation}
{\bf X}\to {\bf X}/ |{\bf X}|^2 \,,
\end{equation}
a minimal surface satisfying $K=0$, is mapped to a surface
satisfying
\begin{equation}
K =  4 {{\bf n}\cdot {\bf X}\over |{\bf X}|^2}\,. \label{eq:KnX}
\end{equation}
This surface will generally possess
a curvature singularity at the origin if this point lies on the surface.
It will also describe an equilibrium as a consequence of
conformal invariance. Thus, if it is also axially symmetric, it
must satisfy Eq.(\ref{eq:quadOmega1}), except perhaps at the origin.
However, this equation can also be cast as
\begin{equation}
{R^2\over 2}\left( C_\perp - C_\|\right)\, K   - c\, {\bf n}\cdot
{\bf X}
 = 0 \,.
\label{eq:quadOmega2}
\end{equation}
Together Eqs. (\ref{eq:KnX}) and (\ref{eq:quadOmega1}) imply
\begin{equation}
{2 R^2\over |{\bf X}|^2} \left( C_\perp - C_\|\right)   = c\,.
\label{eq:quadOmega3}
\end{equation}
The only axially symmetric minimal surfaces are the catenoids, given
parametrically by \footnote{\sf Scaling the catenoid with an inverse
length will give an inverted geometry with the `correct'
dimensions.}
\begin{equation}
R(\Omega) =  R_0^{-1} \, \csc\Omega\,,\quad Z(\Omega) =  R_0^{-1} (
\ln \tan\Omega/2 +\xi_0) \,,
\end{equation}
or, equivalently, as the level set
\begin{equation}
R - R_0^{-1} \cosh (R_0 Z + \xi_0)  =0\,.
\end{equation}
It is characterized by a scale $1/R_0$ and an offset $\xi_0/ R_0$
along the axis. All catenoids are related by scaling and
translation. Inverted in the origin, catenoids map to the surfaces
described by the following transcendental equation
\begin{equation}
{R\over R^2 +Z^2} - {1\over R_0 }\, \cosh \left({R_0 Z\over (R^2
+Z^2)}+ \xi_0\right) =0\,. \label{eq:incat}
\end{equation}
It is simply to check that these surfaces are indeed solutions to
Eq.(\ref{eq:quadOmega1}). Remarkably, each value of the displacement
of the catenoid $\xi_0$ with respect to the point of inversion
describes  a distinct equilibrium.

All of these geometries are compact. The two poles touch because all
distant parts of the original catenoid are mapped to the origin, a
peculiarity of our exact solution. However, a tangent plane does
exist at the origin and one can check that the Euler characteristic
is consistent with a sphere. Their topology, despite appearances, is
spherical.

If $\xi_0=0$, the surface is a symmetric biconcave geometry: a
discocyte. By translating $\xi_0$ along the axis of symmetry, a
physically interesting one-parameter family of solutions is
generated; the up-down symmetry gets broken and there is a
transition from a discocyte to a stomatocyte. Inversion has
connected geometries describing very distinct physical scenarios.
The geometric profiles for three different values of this offset are
illustrated in Fig.(1). The profiles have been rescaled so that they
possess the same surface area.
%The order parameter is the isoperimetric ratio.

\begin{figure}[h]
\begin{center}
\includegraphics[width=5cm]{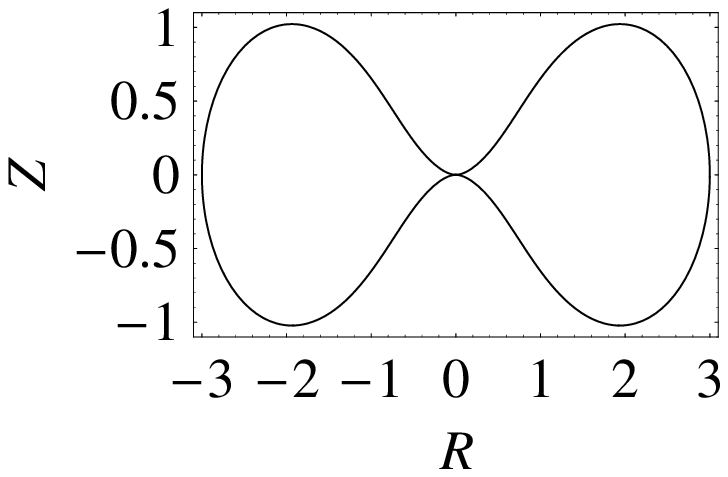}~~~~
\includegraphics[width=5cm]{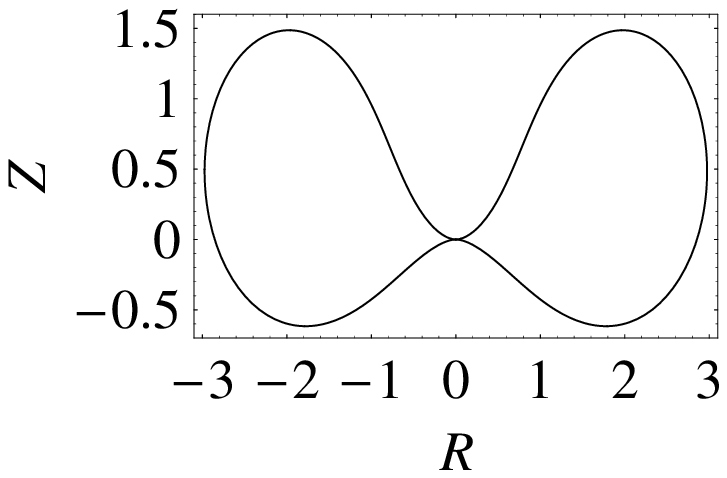}~~~~
\includegraphics[width=5cm]{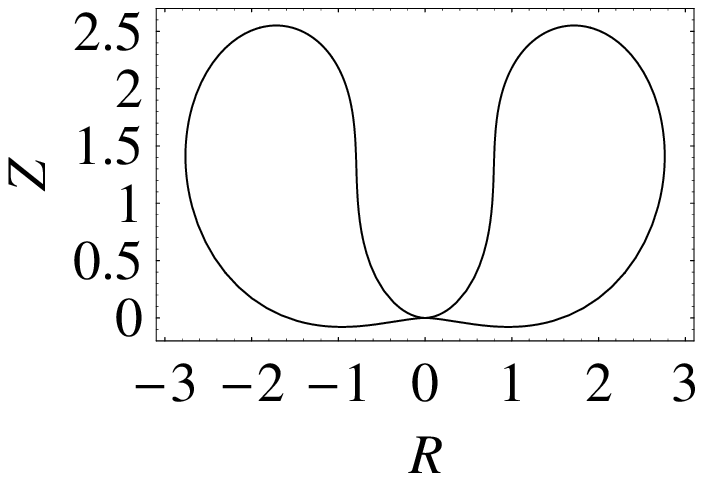}\\
\caption{{\small Geometric profiles for $\xi_0=0,1,2$.}}
\end{center}
\end{figure}

There are superficial similarities with the well-known discocyte to
stomatocyte transition induced by a change in the bilayer asymmetry
%Deuling and Helfrich's pioneering work in the 1970s
(see, for example, \cite{DH,svetina89,seifert}). However, whereas
conformal symmetry is broken globally in that model, here it is
broken only locally (at the poles). The appropriate parameter
describing this transition is the interpolar tension. Its properties
will be discussed in detail in a forthcoming paper. We will thus
limit ourselves to a brief description of the symmetric discocyte
geometry.

Whereas a minimal surface (with $K=0$) is unstressed, as
Eq.(\ref{eq:stress}) demonstrates, the inverted state in under
stress. In the symmetric biconcave geometry, it can be shown that
the two principal curvatures diverge as
\begin{equation}
R_0 C_\|\,\,, R_0 C_\perp  \approx -2 \ln\left( R/R_0\right)
\label{eq:C0}
\end{equation}
The poles are umbilical points of the geometry, albeit singular
ones. It is clear from Eq.(\ref{eq:quadOmega3}) that $C_\perp-C_\|$
will generally be finite.

The curvature singularity at the poles is a manifestation of local
force tethering the poles together. It is unrelated to the fact that
they touch.  The magnitude of this force is given by the Noether
charge. A straightforward calculation gives
\[
c = - {4/R_0}\,.
\]
The sign indicates that the force is directed towards the interior
as we anticipated.

The exact solutions we have written down have their poles joined.
More generally, solutions of Eq.(\ref{eq:quadOmega1}) exist with
poles a fixed distance apart. Like our exact solutions, they will be
singular at these poles.

The tether fixing this distance permits a constraint to be placed on the
isoperimetric ratio: the volume is necessarily deflated below the
maximum spherical value. Remarkably, it is possible to constrain
this ratio without mutilating the conformal invariance of the
energy. This is because the source breaking this symmetry acts only
at isolated points.

It may appear that the solutions we have described with tethered
poles are unphysical. However, even apart from the tethers that are
pulled in the micromanipulation of membranes (see, for example,
\cite{Prost,powers}), it should be emphasized that tethers are
ubiquitous features in biological membranes.  For example, it is
known that the Golgi complex is not an isolated equilibrium
structure. It is stabilized against breakup by a network of
microtubules that are dismantled and reassembled during mitosis. The
tethered membrane we have described is one of those rare tractable
toy models which is relevant in this context.

\section{\sf Conclusions}

A new equation describing the equilibrium of an axially symmetric
fluid membrane has been presented. While our derivation exploits a
specific adapted coordinate system, it is possible to cast the
equation in a form involving only geometrically significant
quantities: the principal curvatures, the support function as well
as the radial distance function. The equation has the remarkable
property that it involves two less derivatives of the embedding functions
than does the corresponding shape equation. This behavior is unexpected.
Even in this simple scenario there is more to be understood.

We have identified new solutions to the axially symmetric shape
equation under the influence of external forces with physical relevance.
This opens a non-perturbative analytical window on processes
that are important in biophysics and
soft matter. Work in this direction is in progress.

Even if one is not interested in the construction of axially
symmetric equilibrium shapes, the problem we have addressed displays
several features of interest from a geometrical point of view. In
particular, the implementation of the variational principle presents
genuine subtleties with a combination of physical constraints that
need to be enforced and unphysical constraints that need to be
relaxed. How to negotiate these points without running into errors
is of broader relevance than the specific problem being addressed.

\vskip3pc \noindent{\Large \sf Acknowledgments}

\vspace{.5cm} We thank Riccardo Capovilla and Markus Deserno for
helpful comments. JG would like to thank IPAM at UCLA for providing
an ideal working environment during spring 2006 where part of this
work was completed. Partial support from CONACyT grants 44974-F and
51111 as well as DGAPA PAPIIT grant IN119206-3 is acknowledged.

\vskip2pc \noindent {\Large \sf APPENDIX}

\vskip3pc \noindent {\Large  \sf Notation}

\vskip1pc A surface is described by three functions ${\bf X} =
(X^1,X^2,X^3)$ of two variables  $\xi^1,\xi^2$. The two coordinate
tangent vectors to the surface are given by ${\bf e}_a= \partial_a
{\bf X}$, $a=1,2$ ($\partial_a = \partial / \partial \xi^a$). Let
${\bf n}$ be the unit normal. The metric tensor induced on the
surface and the extrinsic curvature are then given in terms of these
vectors by $g_{ab}= {\bf e}_a\cdot {\bf e}_{b}$ and $K_{ab}= {\bf
e}_a\cdot \partial_b\, {\bf n}$ \cite{Carmo,Spivak,Ros}. Indices are
raised with the inverse metric. $dA = \sqrt{{\rm det}\, g_{ab}} d^2
\xi$ is the area measure induced on the surface by ${\bf X}$. $K$
denotes (twice) the mean curvature: $K= g^{ab}K_{ab}$

\vskip1pc\noindent {\Large \sf Recovering the shape equation}

\vskip1pc It is simple to confirm  that a first derivative of
equation (\ref{eq:sumelquad}) gives
\begin{equation}
\left(-\partial_{u}+\cos\Omega\right)\left(\partial^{2}_{u}\Omega-
\sin\Omega\cos\Omega\right)-
\frac{1}{2}\left(\left(\partial_{u}\Omega\right)^{2}-\sin^{2}\Omega\right)
\left(\partial_{u}\Omega-\sin\Omega\right)=0\,. \label{eq:elch1}
\end{equation}
A short calculation (substituting Eq.(\ref{curvatures}) into
Eq.(\ref{eq:elch})) confirms that this equation is the axisymmetric
shape equation (\ref{eq:elch}) in isothermal parametrization. It is
also evident that the first term appearing in Eq.(\ref{eq:elch1})
coincides with the 'kinetic' term in (\ref{eq:elch}) whereas the
second term coincides with the cubic potential. This establishes
explicitly the consistency with the reparametrization invariant
shape equation.

\vskip1pc \noindent{\Large \sf  Unphysical tangential stress
associated with constraint on $u$}

\vskip1pc\noindent We will show that a non-vanishing value for the
constant $E$ appearing in the quadrature, associated with an
unphysical constraint on the range of $u$, has its unphysical
counterpart in the stress tensor.
%If we decompose  the stress as ${\bf f}^a = f^{ab}{\bf e}_b + f^a {\bf n}$, o
If this constraint is introduced with a Lagrange multiplier $\tilde
E$, it will add a tangential contribution to the surface stress,
given by
\begin{equation}
{\bf f}^{a}_{{\sf constant}\, u} = -  \tilde E \, {{\bf e}^a \over
R^2}\,. \label{eq:fu}
\end{equation}
This is a surface tension with an unusual inverse $R^{2}$
dependence. If this stress is added to ${\bf f}^a$ consistency with
Eq.(\ref {eq:sumelquad}) requires $E=\tilde E$. The constraint on
$u$ introduces tension. It is not however the usual surface tension.
The energy associated with a constant surface tension $\mu$ is
proportional to area and the corresponding stress is ${\bf
f}^{a}_{\sf tension} = -\mu\, {\bf e}^a $.

The easiest way to derive Eq.(\ref{eq:fu}) is to recast the integral
appearing in Eq.(\ref{eq:1/R}) in the equivalent form,
\begin{equation}
{1\over 2\pi} \int {dA\over R^2}\,. \label{eq:conu}
\end{equation}
Adding such a term breaks the translational invariance of the
energy. Consequently, the associated stress will not be conserved.
The energy remains, however, invariant under translations parallel
to the axis. Thus, the projection of the stress along the axis ${\bf
f}^a\cdot {\bf k}$ is still conserved. Using the auxiliary variables
introduced in \cite{auxil}, it is straightforward to show that the
corresponding addition to the stress is the one given by
Eq.(\ref{eq:fu}). For the record, the conservation law
(\ref{eq:conservf}) is replaced by
\begin{equation}
\nabla_a \,{\bf f}^a = {4 E \over R^4} \,\left ({\bf X} -  ({\bf
X}\cdot {\bf k})\, {\bf k}\right)\,. \label{eq:conservfE}
\end{equation}
This description of stress conservation is invariant under
reparametrization; one need not worry whether the coordinates we
have chosen are well-behaved under variation.

For the same reason as a constraint on $u$ induces an unphysical
stress one also needs to be careful when parametrizing the surface
geometry by arc-length. Instead of constraining the integral
appearing in  (\ref{eq:conu}), one has a constraint on $\int dA/ R$.
The corresponding stress ${\bf f}^{a}_{{\sf constant}\, l}$ is then
proportional to $ {\bf e}^a /R$. This constraint is relaxed in a
manner entirely analogous to that employed for the isothermal
parameter $u$.

Finally, we comment that non-trivial equilibrium single attached
particle configurations do not exist without tension. For, as we
have shown, if $E=0$, then the only solution consistent with regular
geometry at the south pole is part of a sphere.
%In particular, the solutions described in reference \cite{BenSax} appear to be
%artifacts of the constraint on $u$ which introduces the tension
%necessary to support the equilibrium.

\end{document}